\documentclass[aps,onecolumn,preprintnumbers,nofootinbib,superscriptaddress]{revtex4}
\usepackage{amsmath}
\usepackage{graphicx}
\usepackage{amsfonts}
\usepackage{array}
\usepackage{amsthm}
\usepackage{bm}
\usepackage{palatino}
\usepackage{mathpazo}
\usepackage{supertabular}
\usepackage{subfig}
\usepackage[breaklinks]{hyperref}
\usepackage{color}
\usepackage[font=small,labelfont=bf,
justification=justified,
format=plain]{caption}
\usepackage{graphicx}
\usepackage{caption}
\usepackage[font=small,labelfont=bf,justification=justified]{caption}
\usepackage[labelfont=bf,labelsep=quad,justification=justified]{caption}
\usepackage{epstopdf}
\usepackage{physics}
\usepackage[compat=1.1.0]{tikz-feynman}
\usetikzlibrary{calc,decorations.markings,decorations.pathmorphing,
	trees,positioning,arrows}
\providecommand{\keywords}[1]
{
	\small	
	\textbf{\textit{Keywords:}} #1
}
\allowdisplaybreaks
\begin{document}

\title{Quark confinement in presence of both chromoelectric and chromomagnetic fields and the structure of spacetime}

\author{Abdellah Touati}
\email{touati.abph@gmail.com}
\affiliation{Department of physics, Faculty of Exact Sciences, University of Bouira, 10000 Bouira, Algeria}
\author{Soufiane Boukhalfa}
\email{s.boukhalfa@univ-bouira.dz}
\affiliation{Department of physics, Faculty of Exact Sciences, University of Bouira, 10000 Bouira, Algeria}

\begin{abstract}

The strong interaction between quarks inside hadrons in curved spacetime is investigated in the presence of a new non-abelian gauge potential based on the $SU(3)$ group. This potential presented both chromo-electric and chromo-magnetic fields, including a magnetic monopole-like term, together with a radial non-Abelian Coulomb-like component. A spacetime metric induced by the presence of a Yang-Mills field is derived by solving Einstein's equations in the specific limit where the $SU(3)$ gauge symmetry is reduced to an embedded $SU(2)$ subgroup, accompanied by a dynamical $U(1)$ monopole sector. It is explicitly shown that the Schwarzschild radius of the strong interaction between quarks within hadrons corresponds approximately to the size of these latter and that the corresponding, for both quarks and gluons, wave function presents a discontinuity at Schwarzschild surface. The obtained results allow us to interpret the confinement of quarks as a geometric property of spacetime, emerging naturally from its structure without introducing any confinement potential. Moreover, the energy spectra of quarks in presence of chromo-electric and chromo-magnetic fields reproduce the mass of hadrons  with a very good accuracy compared to the experimental data, and the presence of the residual non-abelian term correction enhance our numerical results.

\end{abstract}


\date{\today}
\maketitle

\section{Introduction}

The strong interaction is one of the fundamental forces in nature. Together with the electromagnetic and weak forces, it forms the Standard Model of particle physics, which describes all interactions among particles at the quantum level. Part of the strong force is responsible for binding nucleons (protons and neutrons) within the nucleus, while the other part governs the interaction between quarks that constitute nucleons and, more generally, hadrons. The hypothesis that hadrons are not truly elementary particles was first proposed in 1964 by Murray Gell–Mann \cite{gellman1} and independently by George Zweig \cite{zweig1}, suggesting that they are composed of more fundamental constituents called quarks.

Although this model successfully explains hadron multiplets, it presents a problem: quarks are never observed in isolation, a phenomenon known as quarks confinement. Many models have been proposed to describe quark confinement.  Nonrelativistic examples include the oscillator model \cite{RHD,DF}, the deformed oscillator model \cite{ADR,NI1,NI2,MVNM}, and the Cornell model \cite{EE1,EE2,EE3}.  Relativistic approaches include the Bogoliubov model \cite{PNB}, the MIT bag model \cite{CJJTW,CJJT}, and Dirac equation based potentials of the form \(r^n\), which show good agreement with experimental data \cite{FHZ,RT,LTLT}. Despite their successes in predicting hadron masses accurately, these models are not yet complete descriptions of the strong interaction between quarks inside hadrons.

In this context, a new quantum number called color charge was introduced to better understand the strong interaction \cite{QCD1,QCD2}. Unlike electric charge, color charge comes in three types, but similarly mediates interactions among quarks. This led to the development of quantum chromodynamics (QCD), a non-Abelian gauge theory based on the group $SU(3)$, analogous to quantum electrodynamics (QED) for electric charge with an Abelian gauge group $U(1)$. This theory based on Yang-Mills fields, where the new charge living in the $SU(3)$ group. In QCD, the strong interaction is mediated by massless gluons, which, unlike photons in QED, also carry color charge. This unique feature causes the force between quarks to strengthen as they separate,  explaining why quarks are never seen alone but always confined within hadrons. Various QCD based confinement models have been proposed, such as Refs. \cite{quarkconfinement1,quarkconfinement2}, Coulomb-gauge QCD \cite{quarkconfinement4}, magnetic monopole potentials in non-Abelian gauge theories \cite{monopoles1,fluxtube2,monopoles2}, and flux-tube models \cite{fluxtube1,fluxtube2}.

Recently, a new approach to investigate quark confinement using spacetime structure was proposed by C. C. Barros \cite{Barros3,Barros5}. This approach states that, analogous to how mass curves spacetime in gravitational interactions, the presence of interaction energy from non-gravitational forces such as electromagnetic interactions \cite{Barros1} and strong interactions (including their residual effects) \cite{abdellah1} can also curve spacetime at microscopic scales. In this model, particles are then influenced by the presence of non-gravitational interaction potentials affecting their geodesics  in a purely geometric manner. The framework establishes a robust connection between fundamental interactions and spacetime structure, enabling a unified geometric treatment of these phenomena.

In our work, we operate within the same framework, extending the studies of \cite{Barros3,Barros5,abdellah1} to include non-Abelian gauge fields. This extension enables a geometric investigation of the fundamental strong interaction between quarks inside hadrons, starting from the hypothesis that the interaction energy between color charge directly influences spacetime structure. We propose a non-Abelian gauge potential within the $SU(3)$ group, with chromo-electric and constant chromo-magnetic components. First, we derive the spacetime geometry induced by this non-Abelian gauge potential in the limit where the full \(SU(3)\) symmetry effectively reduces to its embedded \(SU(2)\) Yang-Mills subgroup, supplemented by a dynamical \(U(1)\) magnetic monopole sector. Subsequently, we demonstrate that the phenomenon of quark confinement arises naturally from this geometric framework, without the need for additional confinement potentials. Additionally, we solve the Dirac equation in this curved spacetime and compute masses and sizes of selected hadrons. Our results are compared with experimental data and demonstrate strong agreement.

This paper is organized as follows.  In Sec.~\ref{Sec.2}, we derive the spacetime metric in the presence of chromoelectric and chromomagnetic fields with non-Abelian corrections.  In Sec.~\ref{Sec.3}, we analyze the interaction range, study confinement phenomena, and calculate hadron sizes compared to experimental values.  In Sec.~\ref{Sec.4}, we solve the Dirac equation to obtain hadron masses and compare them with data.  Finally, in Sec.~\ref{Sec.5}, we present our conclusions and remarks.


\section{Spacetime metric in the presence of Yang-Mills field}\label{Sec.2}
In this section, we will derive the metric of the spacetime due to the strong interaction between quarks, within hadrons, in the presence of massless Yang-Mills  field  $\mathcal{A}^a_\mu$ (where $a=1,2,3,..,8$) based on the non-abelian group $SU(3)$. The corresponding strength and energy-momentum tensors are respectively given by 
\begin{align}
	\mathcal{F}^a_{\mu\nu}&=\partial_{\mu}\mathcal{A}^a_\nu-\partial_{\nu}\mathcal{A}^a_\mu+g_{s}f^{abc}A_\mu^bA_\nu^c \label{eq:STYM1}\\
	\mathcal{T}^a_{\mu\nu}&= \mathcal{F}^a_{\mu\sigma}\mathcal{F}^{a\sigma}_{\nu}-\frac{1}{4}g_{\mu\nu}\mathcal{F}^a_{\alpha\beta}\mathcal{F}^{a\alpha\beta}\label{eq:YMT}
\end{align}
where $f^{abc}$ are the structure constants of the $SU(3)$ in Lie algebra, and $g_s$ is the coupling constant of the strong interaction. Similarly to the electric and magnetic fields in the Maxwell theory, one can introduce  these two quantities, known as the chromo-electric field $E^a_i$ and chromo-magnetic field $B^a_i$, defined respectively as
\begin{align}\label{eq:CECM}
	\mathcal{E}^a_i=\mathcal{F}^a_{0i},\qquad \mathcal{B}^a_i=-\frac{1}{2}\epsilon_{ijk} \mathcal{F}^a_{jk}
\end{align}

Similarly to the case of the Einstein-Yang-Mills solution \cite{bizon1,bizon2}, we present here a generalized non-abelian gauge potential in $SU(3)$ group. In this case, we propose the simple non-abelian gauge potential in spherical symmetric, and in the presence of radial non-abelian  Coulomb-likes term together with both chromo-electric and a constant chromo-magnetic (monopole-like term) potentials:

\begin{equation}\label{eq:YMGF}
A_\mu\,dx^\mu
= \phi(r)\,\tau^3\,dt
+ b(\cos\theta\pm1)\,\tau^8  d\varphi
+ w_0\bigl(\tau^4+\sin\theta\,\tau^6+\tau^7\bigr)d\theta
+ \bigg(\frac{d}{r}\,\frac{(\tau^1+\tau^2)}{\sqrt2}+2 S\tau^5 \bigg)\,dr.
\end{equation}
where $\tau^a$ denote the generators of the $SU(3)$ Lie algebra. The constants $\omega_0$ and $b$ characterize the chromo-magnetic components of the gauge potential, while $\phi(r)=\frac{g_s}{4\pi r}$ and $d$ represent the chromo-electric potentials associated with the color charge $g_s$ and a non-abelian Coulomb-like contribution, respectively, where $S$ is a non-abelian radial constant.

In the static case, the non-zero strength tensor components are given by 
\begin{align}
	F_{tr}^1 &= -\frac{g_sd}{r\sqrt2}\,\phi, 
	&F_{tr}^2 &= \frac{g_sd}{r\sqrt2}\,\phi, 
	&F_{tr}^3 &= -\phi',
	&F_{tr}^4 &= -\,g_s\,s\,\phi,\notag
	\\
	F_{t\theta}^5 &= \tfrac12\,g_s\,\phi\,w_0, 
	&F_{t\theta}^6 &= \tfrac12\,g_s\,\phi\,w_0,
	&F_{t\theta}^7 &= -\tfrac12\,g_s\,\phi\,w_0\,\sin\theta,	
	&F_{t\varphi}^a&=0,\notag
	\\
	F_{r\theta}^1 &= -g_sSw_0\sin\theta,
	&F_{r\theta}^2 &= g_sSw_0,
	&F_{r\theta}^3 &= -g_sSw_0,
	&F_{r\theta}^4 &= -\frac{g_sd(1+\sin\theta)}{2r\sqrt2}w_0, \notag
	\\
	F_{r\theta}^5 &= -\frac{g_sd(1-\sin\theta)}{2r\sqrt2}w_0,
	&F_{r\theta}^6 &= \tfrac12\frac{g_sd}{r\sqrt2}w_0,
	&F_{r\theta}^7 &= \tfrac12\frac{g_sd}{r\sqrt2}w_0,
	&F_{r\theta}^8 &= -\sqrt3 g_sSw_0,\notag
	\\
	F_{r\varphi}^4 &=- \sqrt3\,g_s\,S\,b(\cos\theta\pm1),
	&F_{\theta\varphi}^5 &= \tfrac{\sqrt3 g_s\,b w_0}{2}(\cos\theta\pm1)\,, 
	&F_{\theta\varphi}^6 &= +\tfrac{\sqrt3 g_s\,bw_0}{2}(\cos\theta\pm1)\,,
	&F_{\theta\varphi}^8 &= -\,b\sin\theta, 	\notag
	\\
	\nonumber\\
	F_{\theta\varphi}^7 &= +\tfrac{\sqrt3g_s\,b w_0\sin\theta}{2}(\cos\theta\pm1).
\end{align}
Then we can define the total chromo-electric and total chromo-magnetic field invariant by 
\begin{align}
	\mathcal{E}^2=\mathcal{E}_r^2+\mathcal{E}_\theta^2,\quad \mathcal{B}^2=\mathcal{B}_{r\theta}^2 +	\mathcal{B}_{r\varphi}^2+	\mathcal{B}_{\theta\varphi}^2 
\end{align}
where the gauge-invariants can be obtained by summing over the   color index :
\begin{align}
	\mathcal{E}_r^2 &\equiv \sum_a \bigl(F^a_{tr}\bigr)^2
	= \phi'^2 + g_s^2\phi^2\Bigl(\tfrac{d^2}{2r^2}+S^2\Bigr),\\
	\mathcal{E}_\theta^2 &\equiv \sum_a \bigl(F^a_{t\theta}\bigr)^2
	= \tfrac{g_s^2\phi^2w_0^2}{4}\bigl(2+\sin^2\theta\bigr),
\end{align}
and the magnetic components:
\begin{align}
	\mathcal{B}_{r\theta}^2 &\equiv \sum_a \bigl(F^a_{r\theta}\bigr)^2
	= g_s^2S^2w_0^2(5+\sin^2\theta)
	\;+\;\frac{g_s^2d^2w_0^2}{4\,r^2}\Bigl(2+\sin^2\theta\Bigr),\qquad
	\mathcal{B}_{r\varphi}^2 \equiv \sum_a \bigl(F^a_{\theta\phi}\bigr)^2
	= 3\,g_s^2S^2b^2\bigl(\cos\theta\pm1\bigr)^2.\\
	\mathcal{B}_{\theta\varphi}^2 &\equiv \sum_a \bigl(F^a_{\theta\theta}\bigr)^2
	=b^2\sin^2\theta
	\;+\;\tfrac34\,g_s^2b^2w_0^2\bigl(\cos\theta\pm1\bigr)^2(2+\sin^2\theta).
\end{align}

Being inspired in particular by the work of C.C. Barros on Einstein's equations for the electromagnetic interaction \cite{Barros2,abdellah1}, one can write
the Einstein's equations for the Y-M field as
\begin{equation}
	R^{a\mu}_\nu-\frac{1}{2}g^{a\mu}_\nu R^a=\frac{8\pi\alpha_{s}}{m_q^{2}c^{4}}\mathcal{T}^{a\mu}_\nu\label{4.2}
\end{equation}
where the fine structure constant $\alpha$ has been replaced by the coupling constant specific to the strong interaction $\alpha_{s}\equiv g_s^2/4\pi$ ($m_q$ being the quark mass).
A spherically symmetric and static  Schwarzschild-type metric solution of the latter equation  has the following form
\begin{equation}
	ds^{2}=e^{\eta}c^2 dt^{2}-e^{-\eta}dr^{2}-r^{2}d\Omega^{2} \label{4.3}
\end{equation}
where $\eta$ is an arbitrary function of $r$. In this case, the Einstein's equation to be solved is  
\begin{equation}
	\frac{e^{\eta}}{r^{2}}\left(r \eta'+1\right)-\frac{1}{r^{2}}=-K\mathcal{T}^{~0}_{0}\label{4.4}
\end{equation}
where $K=\frac{8\pi\alpha_{s}}{m_q^{2}c^{4}}$. This latter can be put in the following form
\begin{equation}
	\frac{d}{dr}\left(re^{\eta}-r\right)=-Kr^{2}\mathcal{T}^{~0}_{0},\label{4.5}
\end{equation}
where $\mathcal{T}^{~0}_{0}$ is the mixed time component of the energy-momentum tensor. 

In the a stationary state, the energy-momentum tensor reads
\begin{align}
	\mathcal{T}_{00}&=\frac12\biggl[e^{\eta}\,\Bigl(\phi'^2+g_s^2\phi^2\bigl(\tfrac{d^2}{2r^2}+S^2\bigr)\Bigr)
	+\frac{1}{r^2}\,\frac{g_s^2\phi^2w_0^2}{4}\,(2+\sin^2\theta)\biggr]+\frac12\,\biggl[
	\frac{e^{2\eta}}{r^2}\Bigl(g_s^2S^2w_0^2(5+\sin^2\theta)+\frac{g_s^2d^2w_0^2}{4\,r^2}(2+\sin^2\theta)\Bigr)\notag
	\\
	&
	+\frac{e^{2\eta}}{r^2\sin^2\theta}\,\Bigl(3g_s^2S^2b^2(\cos\theta\pm1)^2\Bigr)
	+\frac{e^{\eta}}{r^4\sin^2\theta}\,\Bigl(
	b^2\sin^2\theta
	+\tfrac34\,g_s^2b^2w_0^2(\cos\theta\pm1)^2(2+\sin^2\theta)\Bigr)
	\biggr].\label{4.6}
\end{align}
where  the term  proportional to the strong interaction constant $g_s$ reflects the non-abelian structure of the energy-momentum tensor $\mathcal{T}_{00}$. 
The latter expression can be decomposed into two components, the abelian $(a)$ and the no-abelian $(na)$ ones , as follows 
\begin{equation}
	\mathcal{T}_{00} 
	= \mathcal{T}_{00}^{(\mathrm{a})} \;+\; \mathcal{T}_{00}^{(\mathrm{na})},
\end{equation}
where
\begin{equation}
	\mathcal{T}_{00}^{(\mathrm{a})}
	= \frac{e^{\eta}}{2}\bigg(\,\phi'(r)^2 
	\;+\; \frac{b^2}{\,r^4}\bigg),
\end{equation}
and
\begin{align}	
\mathcal{T}_{00}^{(\mathrm{na})}
	= g_s^2\Bigg[\frac{e^{\eta}}{2}\,\phi^2\bigg(\tfrac{d^2}{r^2}+S^2\bigg)
	\;&+\;\frac{\phi^2w_0^2}{8r^2}\,(2+\sin^2\theta)
	\;+\;\,
	\frac{e^{2\eta}}{2r^2}\Bigl(S^2w_0^2(5+\sin^2\theta)+\frac{d^2w_0^2}{8\,r^2}(2+\sin^2\theta)\Bigr)\notag
	\\
	&
	+\frac{e^{2\eta}}{2r^2\sin^2\theta}\,\Bigl(3S^2b^2(\cos\theta\pm1)^2\Bigr)
	+\frac{e^{\eta}}{2r^4\sin^2\theta}\,\Bigl(
	\tfrac34\,b^2w_0^2(\cos\theta\pm1)^2(2+\sin^2\theta)\Bigr)\Bigg].
\end{align}
The complicated form of this last expression makes it practically impossible to find an analytical solution to Einstein's equations. Hence, and for the sake of simplicity, we eliminate two non-abelian constants part by set $w_0=S=0$, so the non-abelian energy-momentum reduces to
\begin{align}
	\mathcal{T}_{00} 
	= \frac{e^{\eta}}{2}\bigg(\,\phi'(r)^2 
	\;+\; \frac{b^2}{\,r^4}+g_s^2\,\phi^2\tfrac{d^2}{r^2}\bigg).\label{meq}
\end{align}
As one can see, the simplified energy-momentum tensor preserves a residual non-abelian term, which emerges from the commutators between $\tau^3$ and both $\tau^1,$ and $\tau^2$. This new term enables us to observe the effect of the non-abelian correction to the metric that induced by the presence of the non-abelian field \eqref{eq:YMGF} on the quark confinement.

From (\ref{meq}), one can easily determine the expression of  the mixed time component $\mathcal{T}^{~0}_{0}$ given by
\begin{equation}
	\mathcal{T}^{~0}_{0}=\Bigg(\frac{\alpha_s}{8\pi r^4}+\frac{q_m^2}{32\pi^2 r^4}+\frac{\alpha_s^2q_d^2}{32\pi^2 r^4}\Bigg), \label{4.7}
\end{equation}
where $q_m=4\pi\, b$ and $q_d=4\pi\,d$ are the topological magnetic charge \cite{magneticcharge3,bizon1,bizon2,magneticcharge1,magneticcharge2}, and topological non-abelian Coulomb-like charge.

Substituting \eqref{4.7} in \eqref{4.5} and integrating with respect to $r$ the two members of the equation thus obtained, one can obtain
\begin{equation}
	e^{\lambda}=1-\frac{c^2KM(r)}{4\pi r},\label{4.10}
\end{equation}
with
\begin{align}
	M(r)&=\int_{0}^{r}\frac{4\pi}{c^2} r'^{2}\mathcal{T}^{~0}_{0}(r')dr'\notag\\
	&=\int_{0}^{r_0}\frac{4\pi}{c^2} r'^{2}\mathcal{T}^{~0}_{0}(r')dr'+ \int_{r_0}^{r}\frac{4\pi}{c^2} r'^{2}\mathcal{T}^{~0}_{0}(r')dr'\notag\\
	&=M_{0}+\frac{1}{c^2}\int_{r_0}^{r}\Bigg(\frac{\alpha_s}{2 r'^2}+\frac{\alpha_m}{2 r'^2}+\frac{\alpha_s^2\alpha_d}{2r'^2}\Bigg)dr'.
\end{align}
where $\alpha_m=\frac{q^2_m}{4\pi}$ and $\alpha_m=\frac{q^2_m}{4\pi}$ are the new coupling constant specific to the chromo-magnetic field and to the residual non-abelian term respectively. The integration region has been separated into two distinct regions: the interior ($0\leq r'\leq r_0$) and the exterior ($r_0\leq r'\leq r$) of the event horizon. Evaluating the second integral and reporting all this in the equation (\ref{4.10}), then we find, after simplification,
\begin{equation}
	e^{\lambda}=1-\frac{2l}{r}+\frac{\alpha_s(\alpha_s+\alpha_m+\alpha_s^2\alpha_d)}{m_q^2 c^4 r^2},\label{4.11}
\end{equation}
where $l$ is a parameter having the dimension of a length to be fixed hereafter. We  look for a solution of the form
\begin{equation}
	e^{\lambda}=\left[1-\frac{a}{r}\right]^{2}=1-\frac{2a}{r}+\frac{a^{2}}{r^{2}}.\label{4.12}
\end{equation}
By now comparing these last two equations term by term, it comes that
\begin{equation}
	a =l= \frac{\sqrt{\alpha_s(\alpha_s+\alpha_m+\alpha_s^2\alpha_d)}}{m_q c^2},
\end{equation}
and, consequently, the curvature function is defined as
\begin{equation}
	\xi_q(r)\equiv e^{\lambda}=\left[1-\frac{\sqrt{\alpha_s(\alpha_s+\alpha_m+\alpha_s^2\alpha_d)}}{m_qc^{2}r}\right]^{2}, \label{eq:4.13}
\end{equation}
where the term proportional to \(\alpha_d\) represents the non-abelian correction to the metric arising from the Coulomb-like component of the non-abelian gauge potential in the presence of both chromo-electric and chromo-magnetic fields \eqref{eq:YMGF}. This expression therefore captures the residual non-abelian modification of the spacetime geometry due to the embedding of the \(SU(2)\) subgroup into the full \(SU(3)\) gauge group. In the limit of the abelian domination\footnote{The absence of the new non-abelian Coulomb-like term in Eq. \eqref{eq:YMGF}.} we have $\alpha_d=0$, then we find:
\begin{equation}
	\xi_q(r)\equiv e^{\lambda}=\left[1-\frac{\sqrt{\alpha_s(\alpha_s+\alpha_m)}}{m_qc^{2}r}\right]^{2}. \label{eq:4.13'}
\end{equation}
This curvature function describes the spacetime structure in the presence of both a chromo-electric field and a chromo-magnetic (monopole-like) field under the abelian domination, and that means embedding $U(1)$ into the $SU(3)$ group.

Thus, the geometry of spacetime in our case will be described by the following metric
\begin{equation}
	ds^{2}=\xi_q c^2 dt^{2}-\xi_q^{-1}dr^{2}-r^{2}d\Omega^{2}. \label{eq:YM-E-B}
\end{equation}
In absence of the chromo-magnetic field ($\alpha_m=\alpha_d=0$), we retrieve the curvature function found in Refs. \cite{Barros1,Barros3,abdellah1}, i.e.,
\begin{equation}
	\xi(r)=\left(1-\frac{\alpha_{s}}{mc^{2}r}\right)^{2}.\label{4.15}
\end{equation}

\section{Quark confinement and  size of hadrons }\label{Sec.3}

In what follows, we will show how the size of  hadrons can be related to the Schwarzschild radius $r_{s}$ associated with the strong interaction between quarks in  the presence of both chromo-electric and chromo-magnetic fields. The Schwarzschild radius is, obtained by solving the following equation $\xi_q(r_s)=0$, given by
\begin{equation}
	r_{s}=\frac{\sqrt{\alpha_s(\alpha_s+\alpha_m+\alpha_s^2\alpha_d)}}{m_{q}c^{2}} \label{4.26}
\end{equation}
In the case of a quark, inside a nucleon (proton or neutron), the Schwarzschild radius associated with it is approximately $0.831001~$fm, which is of the same order as the radius of the nucleon, which is not negligible given the dimensions of our system (note that the size of a quark is less than $10^{-18}~$m). Classically, the event horizon describes a boundary in a region of spacetime through which  nothing can  pass outward and escape. In the same way, the Schwarzschild radius associated with the strong interaction between quarks delimits a closed surface (a sphere), within the hadrons, where the quarks remain trapped, and whose radius corresponds approximately to the radius of the nucleon by a clever choice of the values of the coupling constants of the strong interaction, i.e. $\alpha_s$ and $\alpha_m$ together with $\alpha_d$. The same values will be used later in the next section to compute the mass of some hadrons (baryons and mesons). Thus, in this approach, the confinement phenomenon appears as an intrinsic property of spacetime without introducing any confinement potential. It should be noticed that if the entire quark content of a hadron is located inside the event horizon, then the classical description of such a system predicts the collapse of all matter into the singularity located at the origin ($r=0$). However, since we are dealing with a quantum system, the uncertainty principle will prevent any collapse. An interesting consequence of this approach is that it is formally impossible for a quark confined inside a hadron to reach the surface $r=r_s$, since the wave function associated with it completely fades out at that point. Indeed, if we solve the radial differential equations \eqref{35} and \eqref{36}, for the curvature function (\ref{eq:4.13}), at the point $r=r_s$, one can easily show that the corresponding wave function at that surface presents a spatial discontinuity, dividing then the space into two completely disconnected regions. This implies that quarks always remain trapped inside hadrons and cannot cross the event horizon under any circumstances. The same conclusions can be drawn for the gluons, bosonic massless particles with spin 1 and carrying a color charge. This type of particles are described by a massless vector field $\mathcal{A}_{\mu}^{a}$, and satisfy the generalized stationary Klein-Gordon equation \eqref{eq2.46} for the curvature function \eqref{eq:4.13}. On the Schwarzschild surface $r=r_s$, we have $\xi_q(r_{s})=0$ and it comes:
\begin{equation}
	\frac{E^{2}}{\hbar^{2}c^{2}}\mathcal{A}_{\mu}^{a}(r_{s})=0\implies \mathcal{A}_{\mu}^{a}(r_{s})=0\label{4.29}
\end{equation}
which shows that the   gluons field function   does indeed have a discontinuity at the event horizon. Consequently, the quarks and gluons are trapped inside the Schwarzschild radius, which describes a boundary region of spacetime beyond which nothing can escape, and that gives a geometric interpretation to the phenomenon of quark confinement inside hadrons.

In the Table \ref{table1} below we give an estimation of the Schwarzschild radius associated with the quarks within some hadrons in terms of the coupling parameter of the strong interaction $\alpha_{s}$ and the new magnetic coupling constants $\alpha_m$ and $\alpha_d$. 
We can observe that our theoretical results predict that the radius of the quarks confinement inside hadrons corresponds approximately to the radius of this hadrons. Moreover, the error from experiment is of the order of $0.00012\%$ to $0.19630\%$ for baryons and $0.0006\%$ to $0.07462\%$ for mesons, which are acceptable results for such a simplistic approach. As an example, take the case of a proton (or neutron) composed of three quarks.  The corresponding Schwarzschild radius is about $0.83058~$fm for a magnetic coupling constant $\alpha_m=0.0012$, and this radius is in the same order of magnitude as the radius of the nucleon itself $(r_{exp}=0.831$fm) \cite{rp2,rp1}, even in the presence of the new non-abelian constant $\alpha_d=0.0016$ where it is about $0.831001~$fm.
Furthermore, we note that in the case of baryons the values of the magnetic constant $\alpha_m$ that give good results are between $0.0012$ and $0.028$. In the case of mesons,  these values are between $0.0024$ and $0.029$. Moreover,  the corresponding values of the new coupling constant, emerged from the non-abelian correction $\alpha_{d}$,  are lying between $0.0001$ and $0.002$ for baryons and between $0.00032$ and $0.00870$ for mesons.  As one can see, for all these values, the new magnetic constant $\alpha_m$ and $\alpha_{d}$ are always less than the coupling constant of the strong interaction $\alpha_{s}$ between quarks inside hadrons (i.e. $\alpha_{d}\ll\alpha_{m}\ll\alpha_{s}$). 

\begin{table}[h]
	\begin{center}
		\begin{tabular}{l l c c c c c c c c c c}
			\hline
			Hadrons &Symbol &$m_{q}$(GeV) & $\alpha_{s}$ & $\alpha_{m}$& $r_{s}$(fm)  \eqref{eq:4.13'} &$r_{s}$(fm)\cite{abdellah1} &$r_{exp}$(fm) & Err (\%) & $\alpha_{d}$ & $r_{s}$(fm) \eqref{eq:4.13}& Err (\%)\\
			\hline
			Proton &$ p ~(uud)$ & 0.380 & 1.60 & 0.0012 & 0.830894  & 0.83058 & 0.831 & 0.01276 & 0.00016  & 0.831001 &  0.00012\\
			
			Sigma & $\Sigma^{-} ~(dds)$ & 0.495 & 1.97 & 0.0280 & 0.790628 & 0.78507 & 0.79372 & 0.38960 & 0.00200 & 0.792162 & 0.19630\\
			
			Xi & $\Xi^{-} ~(dss)$ & 0.550 & 2.00 & 0.0210 & 0.721078 &  0.717322 & 0.72111 & 0.00444 & 0.00010 & 0.721149 & 0.00541\\
			
			Pion & $\pi^{-}~ (d\bar{u})$ & 0.071 & 0.24 & 0.0024 & 0.670132 & 0.66681 & 0.67082 & 0.10300 & 0.00870 & 0.670824 & 0.00060\\
			
			Kaon & $K^{-} ~(s\bar{u})$ & 0.280 & 0.87 & 0.0040 & 0.614333 &  0.61293 & 0.61644 & 0.34180 & 0.00600 & 0.615980 & 0.07462 \\
			
			Rho & $\rho^{-}~ (u\bar{d})$ & 0.477 & 1.80 & 0.0120 & 0.746868 & 0.74439 & 0.74833 & 0.19536 & 0.00220 & 0.748335 & 0.00067\\
			
			J/Psi & $J/\Psi~(c\bar{c})$ & 1.79 & 1.00 & 0.0200 & 0.111300 &  0.11130 & -  & -  & 0.00032 & 0.111317 & -\\
			
			Upsilon & $\Upsilon~(b\bar{b})$ & 5.50 & 1.05 & 0.0290 & 0.038176 &  0.03766 & - & - & 0.003100 & 0.038182 & -\\
			\hline
		\end{tabular}
		\caption{Radius of few hadrons (baryons and mesons), obtained using the relation (\ref{4.26}), in comparison to experimental data \cite{rp2,rp1,Kaon,Rho}.}
		\label{table1}
	\end{center}
\end{table}

\section{Energy spectrum and  mass of hadrons}\label{Sec.4}
In order to obtain the energy spectrum of the interaction between quarks in presence of chromo-electric and chromo-magnetic fields and the non-abelian correction, we need to solve the Dirac radial equations \eqref{35} and \eqref{36} for the curvature function \eqref{eq:4.13}. 
The energy spectrum  can be directly deduced following the same steps as in Refs. \cite{Barros3,abdellah1}. After some algebra, one can easily show that
\begin{equation}
	E_n= m_{q}\left[\frac{1}{2}-\frac{n^{2}}{8\alpha_s(\alpha_s+\alpha_m+\alpha_s^2\alpha_d)}+\frac{n}{4\sqrt{\alpha_s(\alpha_s+\alpha_m+\alpha_s^2\alpha_d)}}
	\sqrt{\frac{n^{2}}{4\alpha_s(\alpha_s+\alpha_m+\alpha_s^2\alpha_d)}+2}\right]^{1/2},\,\text{with}\quad n\geq1. \label{eq:4.30}
\end{equation}

\begin{figure}[h]
	\centering
	\includegraphics[width=0.24\textwidth]{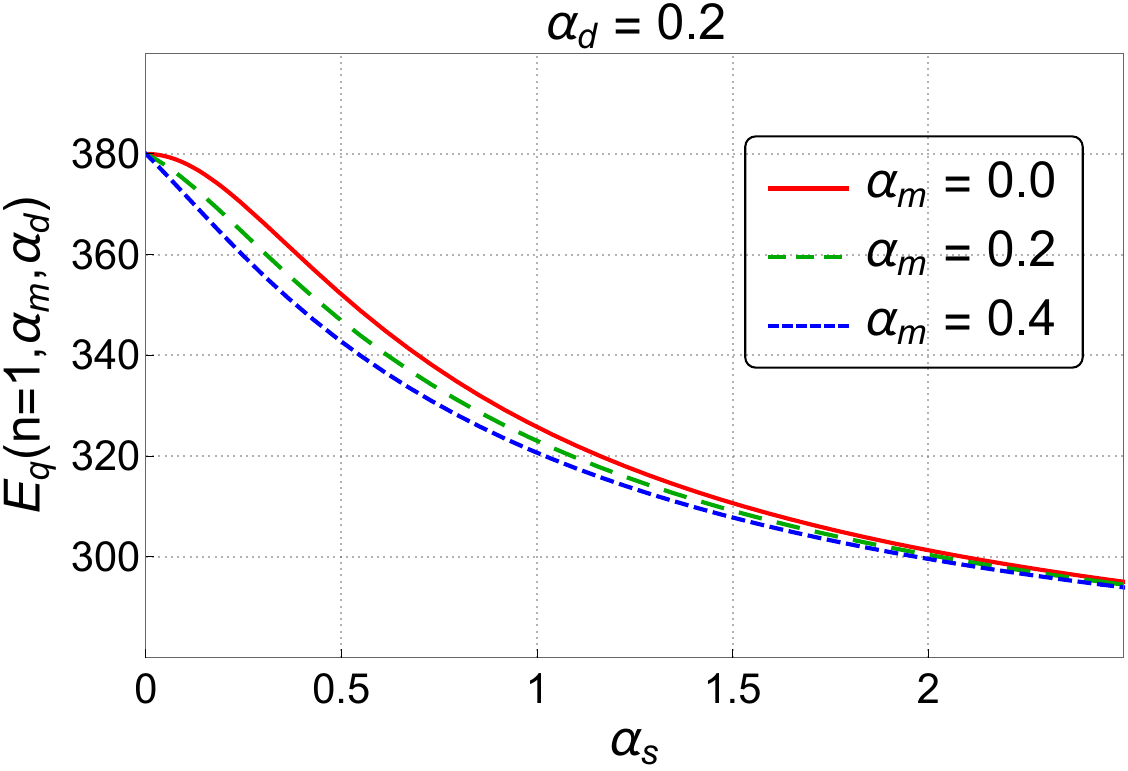}
	\includegraphics[width=0.24\textwidth]{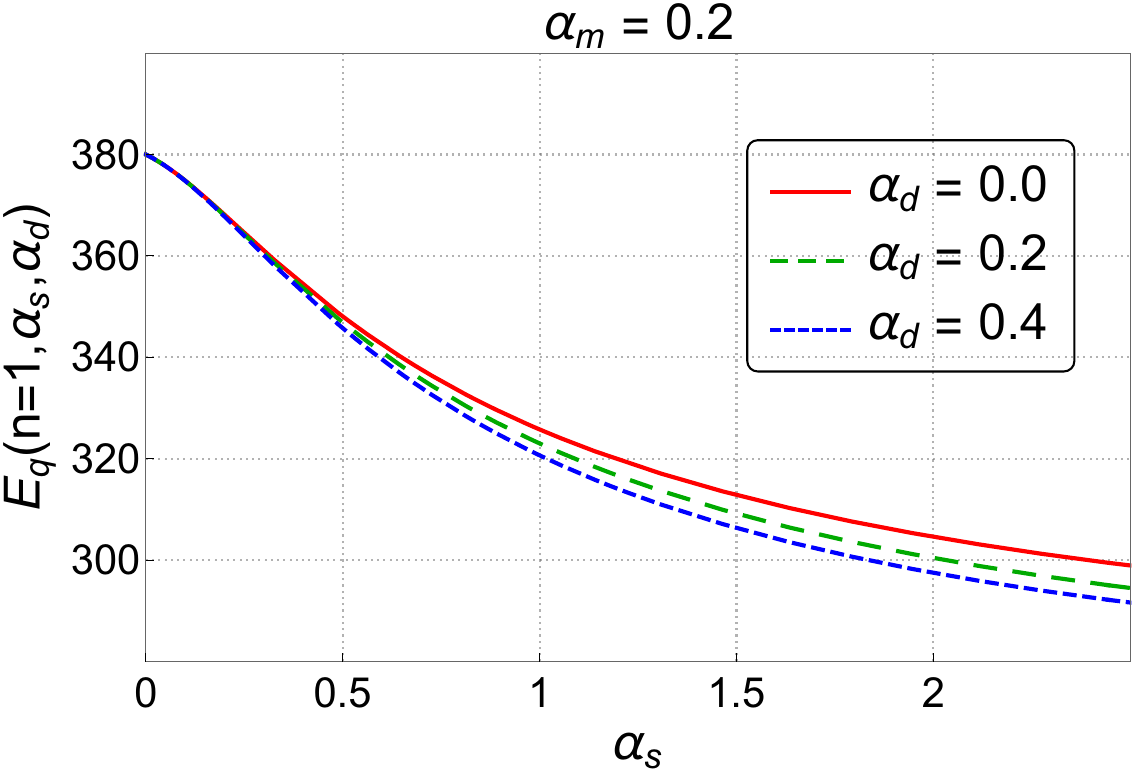}
	\includegraphics[width=0.24\textwidth]{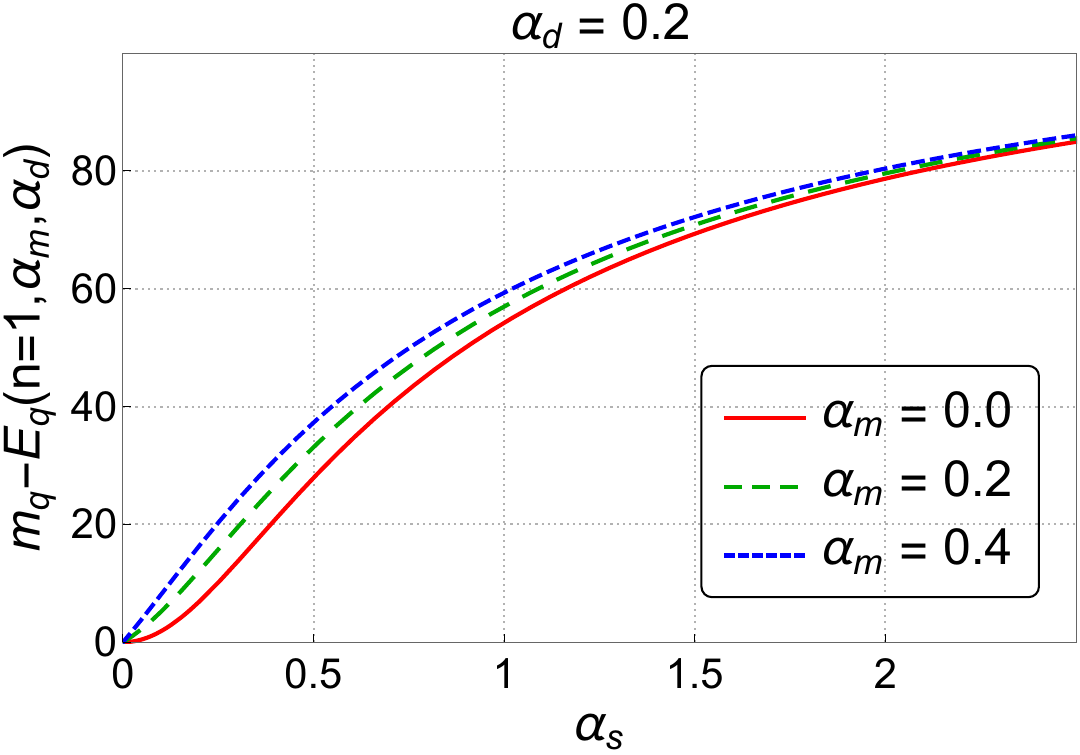}
	\includegraphics[width=0.24\textwidth]{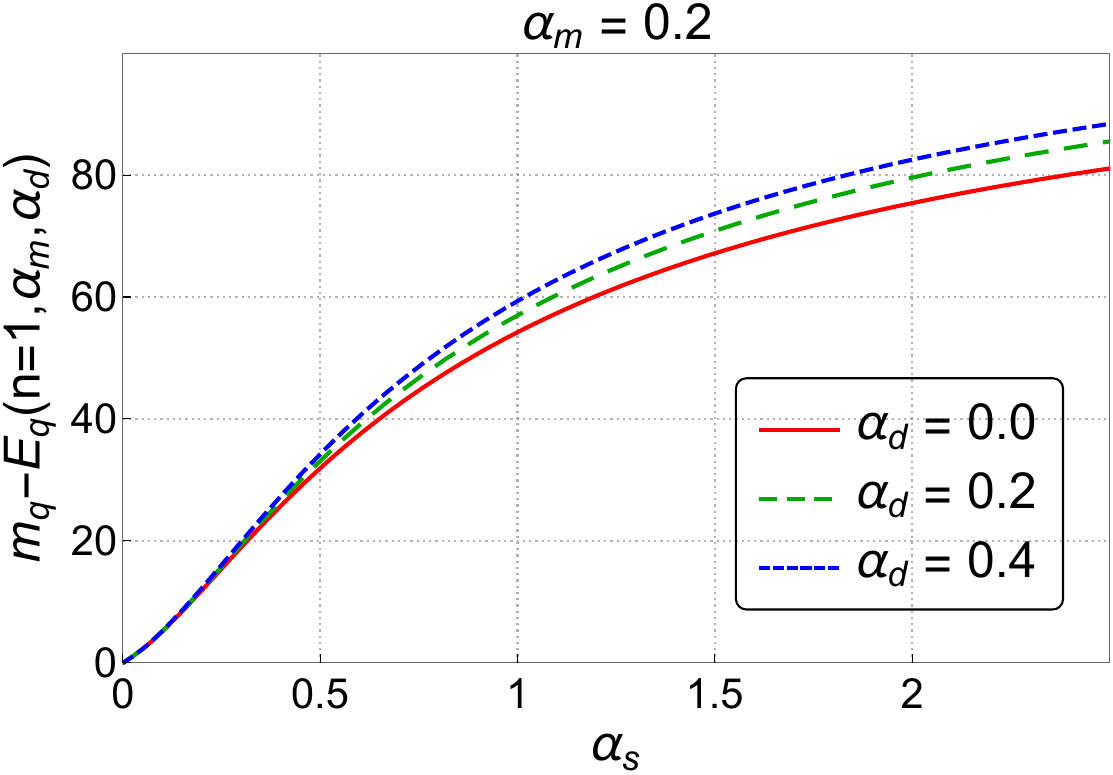}
	\caption{\textbf{Left side:} Variation of the fundamental energy of a quark within the hadron in terms of the strong coupling constant, $\alpha_{s}$, for different magnetic constants $\alpha_m$ and $\alpha_{d}$ respectively. \textbf{Right side:} Variation of the interaction energy of a quark within the hadron in terms of the strong coupling constant, $\alpha_{s}$, for different magnetic constants $\alpha_m$ and $\alpha_{d}$ respectively.}\label{fig:1}
\end{figure}
In Fig.~\ref{fig:1} we show the variation of both the quark’s fundamental energy and its interaction energy inside a hadron as functions of the strong coupling constant~\(\alpha_s\), for different values of the magnetic constants \(\alpha_{m}\) and \(\alpha_{d}\) in the presence of the non-abelian correction. On the left two panels, one sees that the effect of the new magnetic constant \(\alpha_{m}\) is significant at small coupling constant \(\alpha_s\) but becomes negligible at large \(\alpha_s\). Conversely, the non-abelian Coulomb-like constant \(\alpha_{d}\) has a negligible impact at small \(\alpha_s\) and becomes important at large \(\alpha_s\); like \(\alpha_{m}\), it reduces the quark’s fundamental energy in the same manner as increasing \(\alpha_s\). On the right two panels, we plot the interaction energy of the quark inside the hadron. Here, the energy lost by the quark on the left panels is transferred into interaction energy, which grows with \(\alpha_s\). Increasing either magnetic constant, \(\alpha_{m}\) or \(\alpha_{d}\), also raises the interaction energy. Again, the effect of \(\alpha_{m}\) is confined to the low \(\alpha_s\) region, while \(\alpha_{d}\) dominates at high \(\alpha_s\).

In the limit of the abelian domination $\alpha_{d}=0$, the above expression reduced to the following one:
\begin{equation}
	E_n= m_{q}\left[\frac{1}{2}-\frac{n^{2}}{8\alpha_s(\alpha_s+\alpha_m)}+\frac{n}{4\sqrt{\alpha_s(\alpha_s+\alpha_m)}}
	\sqrt{\frac{n^{2}}{4\alpha_s(\alpha_s+\alpha_m)}+2}\right]^{1/2},\,\text{with}\quad n\geq1. \label{eq:4.30'}
\end{equation}
which describes the energy of quarks inside a hadron in the presence of a like-Coulomb color electric field and a constant color magnetic field under abelian domination (i.e.\ embedding \(U(1)\) into \(SU(3)\)). In the special case \(\alpha_m=\alpha_d=0\), this formula further reduces to the result obtained in Ref.~\cite{Barros3}.

The mass of a hadron (expressed in MeV), with several quarks inside, can be evaluated by the following relation \cite{Barros3}:
\begin{equation}
	M_{\text{hadron}}=\sum_{i=1}^{l} E_{1,i}(m_q) \label{4.31}
\end{equation}
where $l$ and $m_q$  represent respectively the quarks number and the effective mass of the quark inside the hadron. The effective mass of a quark inside a proton according to some papers \cite{QM1,QM4} is about $380~$Mev, much greater than the free mass of a $u$ or $d$ quark (the constituents of a proton) which is about few electron-volts ($2$ to $5$ eV). According to Einstein's mass-energy equivalence relation, the interaction energy of the surrounding self-consistent strong field that surrounds the quark is converted into a mass, which explains where the effective mass of a quark comes from. Thus, for a proton consisting of three quarks, the relation (\ref{4.31}) predicts a mass of about $938.271~$MeV (with $\alpha_{s}=1.6$ and together with both magnetic constant $\alpha_m=0.0012$ and non-abelian constant $\alpha_{d}=0.00016$), which is very close to the experimental mass of a proton, which is about $938.27~$MeV.

In the Table \ref{table2} below, we present an estimation of some hadrons masses using the coupling parameter $\alpha_{s}$ and the both of magnetic constant $\alpha_{m}$ and non-abelian one $\alpha_d$ together with the effective masses of the quarks ($m_{q}$). Their masses $M_{exp}$ \cite{PpN1} and their radii $r_{exp}$ measured experimentally are also reported in Refs.  \cite{rp2,rp1,Kaon,Rho}. On this one, we observe that the theoretical results predicted for the masses of the hadrons (baryons and mesons) are very close to the experimental results with sometimes a precision of the order of a $1/100000$th of GeV. Indeed, the deviation from experiment is of the order of $0.00011\%$ to $0.46304\%$ for baryons and $0.00011\%$ to $0.71505\%$ for mesons, which are more than acceptable results for such a simplistic and heuristic approach.
Moreover, we note that, in the case of baryons, the values of the coupling constant that give good results are between $1.5$ and $2$ (which are reasonable values to describe the interaction of quarks inside baryons), as well as, in the case of mesons,
these are between $0.2$ and $1.4$, with the exception of the meson $\rho$ which
has a slightly higher coupling constant ($\alpha_{s}=1.8$). We also note that the masses of the baryons are proportional to the value of the coupling constant, while their radii are inversely proportional to it. Therefore, the mass of a baryon depends directly on the effective mass of the quarks that compose it. However, this same mass already takes into account the interaction energy of the surrounding gluonic field that surrounds the quarks inside a baryon. Consequently, the greater the effective mass of the quarks, the stronger the coupling between them inside a baryon and the more inertia it will have.
On the other hand, the greater the effective mass of the quarks inside a baryon, the smaller the confinement radius associated with them, and therefore the smaller the radius of the baryon in question. We also observe, in the case of mesons, the same behavior of the coupling constant in relation to their masses and radii, with the exception of the $\rho$ meson which deviates slightly from this rule. This can be attributed to the choice of a spherically symmetric metric in the description of all hadrons (baryons and mesons), while another choice of symmetry (e.g. an elliptically symmetric metric) may be proved more fruitful and will probably improve the description of some mesons.
\begin{table}[h]
	\begin{center}
		\begin{tabular}{l l c c c c c c c c c c}
			\hline
			Hadrons &Symbol &$m_{q}$(GeV) & $\alpha_{s}$& $\alpha_{m}$ & $M$(GeV) \eqref{eq:4.30'}  &$M$(GeV)\cite{abdellah1} &$M_{exp}$(GeV) & Err (\%)  & $\alpha_{d}$ & $M$(GeV) \eqref{eq:4.30} & Err (\%)\\
			\hline
			Proton &$ p ~(uud)$ &0.380 & 1.60 & 0.0012 & 0.938284 & 0.93832 & 0.93827 & 0.00053  & 0.00016 & 0.938271 & 0.00011\\
			
			Sigma & $\Sigma^{-} ~(dds)$ & 0.495 & 1.97 & 0.0280 & 1.19656 & 1.19736 & 1.19745 & 0.07430 & 0.00200 & 1.196340 & 0.09260\\
			
			Xi & $\Xi^{-} ~(dss)$ & 0.550 & 2.00 & 0.0210 & 1.32784 & 1.32850 & 1.32171 & 0.46379    & 0.00010 & 1.327830 & 0.46304\\
			
			Pion & $\pi^{-}~ (d\bar{u})$ & 0.071 & 0.24 & 0.0024 & 0.138578 & 0.13861 & 0.13957 & 0.71075  & 0.00870 & 0.138572 & 0.71505\\
			
			Kaon & $K^{-} ~(s\bar{u})$ & 0.280 & 0.87 & 0.0040 & 0.492392 & 0.49252 & 0.49368 & 0.26089    & 0.00600 & 0.492252 & 0.28926\\
			
			Rho & $\rho^{-}~ (u\bar{d})$ & 0.477 & 1.80 & 0.0120 & 0.775713 & 0.77597 & 0.77526 & 0.05843  & 0.00220 & 0.775563 & 0.03908\\
			
			J/Psi & $J/\Psi~(c\bar{c})$ & 1.79 & 1.00 & 0.0200 & 3.09696 & 3.10037 & 3.09690 & 0.00194   & 0.00032 & 3.096910 & 0.00032\\
			
			Upsilon & $\Upsilon~(b\bar{b})$ & 5.50 & 1.05 & 0.0290 & 9.46049 & 9.47479 & 9.46030 & 0.00201  & 0.00031 & 9.460310 & 0.00011\\
			\hline
		\end{tabular}
		\caption{Masses of some hadrons (baryons and mesons), obtained using the relation (\ref{4.31}), in comparison to their  experimental values.}
		\label{table2}
	\end{center}
\end{table}


\section{Conclusion}\label{Sec.5}

In this work, we have investigated the strong interaction between quarks and the confinement phenomenon inside hadrons by introducing a new non-abelian gauge potential in the \(SU(3)\) group. We included both a chromo-electric field and a constant chromo-magnetic (monopole-like) field, supplemented by a radial non-abelian Coulomb-like term. In the special limit where only the embedded \(SU(2)\) subgroup remains active, together with a dynamical \(U(1)\) monopole sector, and our construction reduces to the familiar \(SU(2)\) Yang–Mills scenario enriched by an explicit abelian magnetic monopole-like contribution.

First, we derived the new metric in terms of the strong coupling and the magnetic constants together with a residual non-abelian correction term, and showed that this one leads naturally  to the emergence of the quarks and gluons confinement inside hadrons without introducing any confinement potential. Then we estimated the corresponding Schwarzschild radius for the basic strong interaction and showed that its value is approximately equal to the experimental size of the hadron, and the presence of the magnetic constant predicted a good results comparing to the experimental ones, and sometimes with an error of about $0.00444\%$, while in the presence of the residual non-abelian term this error reduces to $0.00012\%$, which is a good results for such a oversimplified approach. 

Secondly, we solved the Dirac equation to obtain the energy spectrum of the interaction between quarks and estimate some hadrons masses. 
The results thus obtained are, for the most part, very close to the experimental data and sometimes with an error of $0.00053\%$ in the presence of the magnetic coupling constant, 
which reduces to $0.00011\%$ when both magnetic and residual non-abelian term are present.

At the end, based on the obtained results one can conclude that the present approach enables us to explain the phenomenon of color confinement within hadrons in a simple and elegant way as an inherent property of spacetime, and to estimate the radii and the masses of some hadrons with a very a good accuracy comparing to experimental data.

\appendix

\section{Tetrad formalism}\label{sec:A}

Considering now a general form of the Schwarzschild type metric which is written as a function of the metric $g_{\mu\nu}$, and with Einstein's notation of summation:
\begin{equation}
	ds^{2}=g_{\mu\nu}dx^{\mu}dx^{\nu}\label{2}
\end{equation}
The covariant and contravariant components of the metric $g_{\mu\nu}$:
\begin{equation}
	[g_{\mu\nu}]=\left(\begin{array}{cccc}
		e^{2f}	& 0 & 0 & 0 \\ 
		0	& -e^{2g} & 0 & 0 \\ 
		0	& 0 & -r^{2} & 0 \\ 
		0	& 0 & 0 & -r^{2}sin^{2}\theta 
	\end{array} 
	\right)\label{3}
\end{equation}
and:
\begin{equation}
	[g^{\mu\nu}]=\left(\begin{array}{cccc}
		e^{-2f}	& 0 & 0 & 0 \\ 
		0	& -e^{-2g} & 0 & 0 \\ 
		0	& 0 & -\frac{1}{r^{2}} & 0 \\ 
		0	& 0 & 0 & -\frac{1}{r^{2}sin^{2}\theta} 
	\end{array} 
	\right)\label{4}
\end{equation}
We take the following notation $e^{\mu}_{a}$ for tetrads, with $\mu$ representing the general coordinates (curved spacetime) and the index $a$ denoting the local coordinates (flat spacetime), and which are defined by:
\begin{equation}
	e^{\mu}_{a}=\frac{\partial x^{\mu}}{\partial y^{a}},\qquad e^{a}_{\mu}=(e^{\mu}_{a})^{-1}=\frac{\partial y^{a}}{\partial x^{\mu}} \label{5}
\end{equation}
These tetrads relate the curved spacetime metric to the flat spacetime metric by the following diagonalization relation:
\begin{equation}
	e^{\mu}_{a}e^{\nu}_{b}g_{\mu\nu}=\eta_{ab},\qquad e^{a}_{\mu}e^{b}_{\nu}\eta_{ab}=g_{\mu\nu}\label{6}
\end{equation}
we take the following signature for the Minkowski metric $\eta_{ab}=diag(1,-1,-1,-1)$. Or the covariant derivative (for spinors) $\nabla_{\mu}$ is given by:
\begin{equation}
	\nabla_{\mu}=\partial_{\mu}+\Gamma_{\mu}\label{7}
\end{equation}
with $\Gamma_{\mu}$ called the spin connection, and which are defined by:
\begin{equation}
	\Gamma_{\mu}=\omega_{ab\mu}[\gamma^{a},\gamma^{b}]\label{8}
\end{equation}
with $\gamma^{a}$ being the Dirac matrices in flat spacetime and which satisfy the anticommutation relation $[\gamma^{a},\gamma^{b}]=2\eta^{ab}$, the components of $\omega_{ab\mu}$ are calculated by:
\begin{equation}
	\omega_{ab\mu}=\eta_{ac}\omega^{c}_{b\mu}=\eta_{ac}\left[e^{c}_{\mu}\partial_{\nu}(e^{\mu}_{b})+e^{c}_{\mu}e^{\sigma}_{b}\Gamma^{\mu}_{\sigma\nu}\right]\label{9}
\end{equation}
the tensor $\omega_{ab\mu}$ is an anti-symmetric tensor in the first two indices $\omega_{ab\mu}=-\omega_{ba\mu}$ where the symbols $\Gamma^{\mu}_{\sigma\nu}$ are the Christoffel symbols and are defined by (for the Schwarzschild metric you can see the reference ...):
\begin{equation} 
	\Gamma^{\mu}_{\sigma\nu}=\frac{1}{2}g^{\mu\rho}\left[\partial_{\sigma} g_{\rho\nu}+\partial_{\nu} g_{\sigma\rho}-\partial_{\rho} g_{\sigma\nu}\right]\label{10}
\end{equation} 
	We use the relations (\ref{6}) and (\ref{3}-\ref{4}), we can calculate the components of the tetrads:
\begin{align} e^{\mu}_{a}&=diag\left(e^{-f},e^{-g},\frac{1}{r},\frac{1}{rsin\theta}\right)\label{11},\\ e^{a}_{\mu}&=diag\left(e^{f},e^{g},r,rsin\theta\right)\label{12} 
\end{align}
We calculate the components $\Gamma^{\mu}_{\sigma\nu}$ and $\omega_{ab\mu}$ and we replace them in the relation (\ref{8}), we find: 
\begin{align} \Gamma_{0}=\frac{f'}{4 }&e^{f-g}[\gamma^{0},\gamma^{1}],\quad\Gamma_{1}=0,\quad\Gamma_{2}=\frac{e^{-g}}{4}[\gamma^{1},\gamma^{2}]\notag\\
&\Gamma_{3}=\frac{sin\theta}{4}e^{-g}[\gamma^{1},\gamma^{2}]+\frac{cos\theta}{4}[\gamma^{2},\gamma^{3}] \label{13}
\end{align}
In this work we use the diagonal gauge of the Dirac matrix:
\begin{equation}
	\gamma^{0}=\beta, \gamma^{i}=\beta\alpha^{i}\label{14}
\end{equation}
with:
\begin{equation}
	\beta=\left(\begin{array}{cc}
		1&0\\
		0&-1	
	\end{array}
	\right),\quad
	\gamma^{i}=\left(\begin{array}{cc}
		0&\sigma^{i}\\
		-\sigma^{i}&0
	\end{array}
	\right),\quad
	\alpha^{i}=\left(\begin{array}{cc}
		0&\sigma^{i}\\
		\sigma^{i}&0
	\end{array}
	\right)
	\label{15}
\end{equation}
where $\sigma^{i}$ are the $2\times2$ Pauli matrices.

\subsection{Dirac equation in curved spacetime}\label{sec:B}
In this subsection, we derive the Dirac equation in the curved spacetime for a general form of a Schwarzschild-like metric \eqref{3}.
The Dirac equation in curved spacetime, with $\hbar=c=1$ is written as follow:
\begin{equation}
	(i\gamma^{\mu}\nabla_{\mu}-m)\Psi(t,r)=0
\end{equation}
where $\nabla_{\mu}$ is the covariant derivative and the Dirac matrices in general coordinates are connected with the Dirac matrices in local coordinates by the following relation $\gamma^{\mu}=e^{\mu}_{a}\gamma^{a}$, using the relations (\ref{7}-\ref{11}-\ref{12}-\ref{13}), the Dirac equation is written as:
\begin{align}
	\left[ie^{0}_{0}\gamma^{0}(\partial_{0}+\Gamma_{0})+ie^{1}_{1}\gamma^{1}\right.\left.(\partial_{1}+\Gamma_{1})+ie^{2}_{2}\gamma^{2}(\partial_{2}+\Gamma_{2})+ie^{3}_{3}\gamma^{3}(\partial_{3}+\Gamma_{3})-m\right]\Psi(t,r)=0
\end{align}
In order to replace the values of the spin connections with the Dirac matrices (\ref{14}-\ref{15}) and to rearrange the terms, the Dirac equation is written as:

\begin{align} 
	\left[-ie^{f-g}\alpha^{1}\left(\partial_{r}+\frac{1}{r}+\frac{f'}{2}\right)-i\alpha^{2}\frac{e^{f}}{r}\left(\partial_{\theta}+\frac{cot\theta}{2}\right)-i\frac{e^{f}}{ r sin\theta}\alpha^{3}\partial_{\phi}+\beta e^{f}m\right]\Psi(t,r)=i\frac{\partial \Psi(t,r)}{\partial t} 
\end{align} For the stationary case: 
\begin{equation} 
	\Psi(t,r)=e^{-iEt}\psi(r),
\end{equation} 
Then we apply the unitary transformation on the spinor and on the Hamiltonian $\psi'=U\psi$ and $H'=UHU^{\dagger}$ with $U=\frac{1}{2}(1_{4}+\gamma^{2}\gamma^{1}+\gamma^{1}\gamma^{3}+\gamma^{3}\gamma^{2})$, so the Dirac equation will become: 
\begin{align} \left[-ie^{f-g}\alpha^{3}\left(\partial_{r}+\frac{1}{r}+\frac{f'}{2}\right)-i\frac{e^{f}}{r}\alpha^{1}\left(\partial_{\theta}+\frac{cot\theta}{2}\right) c{e^{f}}{r}\frac{\alpha^{2}}{sin\theta}\partial_{\phi}+\beta e^{f}m\right]\psi'=E\psi'
\end{align} 
We set:
\begin{equation}
	\hat{S}=1+\sigma^{1}\left(\partial_{\theta}+\frac{cot\theta}{2}\right)+\frac{\sigma^{2}}{\sin\theta}\partial_{\phi}
\end{equation}
our equation is now written:
\begin{equation}
	\left[-i\alpha^{3} e^{f-g}\left(\partial_{r}+\frac{1}{r}+\frac{f'}{2}\right)-i\alpha\frac{e^{f}}{r}(\hat{S}-1)+\beta e^{f}m\right]\psi'=E\psi'
\end{equation}
We write the spinor $\psi'$ in two components with the separation of the variables:
\begin{equation} 
	\psi'(r,\theta,\phi)=\left(\begin{array}{c} \frac{F(r)}{r}e^{-f/2}\chi(\theta,\phi) \\ -i\sigma^{3} \frac{G(r)}{r}e^{-f/2}\chi(\theta,\phi) \end{array} \right)
\end{equation}
where $\chi$ are the spherical spinors. Using the above decomposition and with some algebra we find the final form with $e^{f/2}=\sqrt{\xi(r)}$:
\begin{align}
	\sqrt{\xi}\frac{\partial F(r)}{\partial r}+(1+k)\frac{F(r)}{r}&=(\frac{E}{\sqrt{\xi}}+m)G(r)\label{35}\\
	\sqrt{\xi}\frac{\partial G(r)}{\partial r}+(1-k)\frac{G(r)}{r}&=-(\frac{E}{\sqrt{\xi}}-m)F(r)\label{36}
\end{align}
These equations are the same one that obtained in Ref. \cite{Barros1} when we use \eqref{4.15}.

\section{Klein-Gordon equation in curved spacetime}\label{sec:C}
In a curved spacetime, equipped with a Schwarzschild-type metric, we can establish a quantum wave equation, for a massive particle of spin 0, for the energy operator is given by (for more detail see \cite{Barros1})
\begin{equation}
	E=i\hbar\partial_{t}^{Sch}= \frac{i\hbar}{\sqrt{\xi}}\frac{\partial}{\partial t}\label{eq2.15}
\end{equation}
and momentum operator
\begin{equation}
	\vec{p}^{2}_{Sch}=-\hbar^{2}\left[\frac{\sqrt{\xi}}{r^{2}}\frac{\partial}{\partial r}\left(r^{2}\sqrt{\xi}
	\frac{\partial}{\partial r}\right)- \frac{\vec{L}^2}{r^2}\right]\label{eq2.14}
\end{equation}
where $\vec{L}^2$ is the squared modulus of the orbital angular momentum operator \cite{Barros1}.
Then put the above quantities into the dispersion equation 

\begin{equation}
	\frac{E^{2}}{\xi}=\vec{p}_{Sch}^2 c^2 + m^{2}c^{4} \label{eq2.34}
\end{equation}
Now making the obtained relation act on a wave function $\Psi(\vec{r},t)$. We then obtain the generalized Klein-Gordon equation,
following:
\begin{align}
	-\frac{\hbar^{2}}{\xi^{2}}\frac{\partial^{2}\Psi(\vec{r},t)}{\partial t^{2}}&=
	-\hbar^{2}c^{2}\vec{\nabla}_{Sch}^2 \Psi(\vec{r},t)+m^{2}c^{4}\Psi(\vec{r},t)
\end{align}
taking into account (\ref{eq2.14}),
\begin{equation}
	-\frac{\hbar^{2}}{\xi^{2}}\frac{\partial^{2}\Psi(\vec{r},t)}{\partial t^{2}}=
	\left\{-\hbar^{2}c^{2}\left[\frac{\sqrt{\xi}}{r^{2}}\frac{\partial}{\partial r}\left(r^{2}\sqrt{\xi}\frac{\partial}{\partial r}\right)
	-\frac{\vec{L}^2}{r^{2}}\right] +m^{2}c^{4}\right\}\Psi(\vec{r},t) \label{eq2.43}
\end{equation}
In the stationary case and taking into account the spherical symmetry of the problem,
we can seek a solution of the form:
\begin{equation}
	\Psi(\vec{r},t)\equiv\Psi(r,\theta,\phi,t)= U(r)Y^{m}_{l}(\theta,\phi)\exp\left(-i\frac{Et}{\hbar}\right)\label{eq2.44}
\end{equation}
where $U(r)$ is a radial function and $Y^{m}_{l}(\theta,\phi)$ are the spherical harmonics, eigenfunctions of the orbital angular momentum square operator satisfying the following eigenvalue equation:
\begin{equation}
	\vec{L}^2 Y^{m}_{l}(\theta,\phi)= \hbar^2 l(l+1)Y^{m}_{l}(\theta,\phi)\label{eq2.45}
\end{equation}
where $l$ and $m$ are the orbital and magnetic quantum numbers respectively. Now reporting the solution (\ref{eq2.44})
in (\ref{eq2.43}), and taking into account (\ref{eq2.45}),
we obtain, after simplification and rearrangement of the different terms, the differential equation,
which must satisfy the radial part of the wave function, which follows:
\begin{equation}
	\frac{\sqrt{\xi}}{r^{2}}\frac{d}{dr}\left(r^{2}\sqrt{\xi}\frac{d}{dr}\right)U(r)
	+\left(\frac{E^{2}}{\hbar^{2}c^{2}\xi^{2}}-\frac{m^{2}c^{2}}{\hbar^{2}}-\frac{l(l+1)}{r^{2}}\right)U(r)=0\label{eq2.46}
\end{equation}
Note that to completely solve this equation, it is necessary to know the explicit form of the function $\xi(r)$ which depends
on the interaction potential $V(r)$ relative to the system studied.


\bibliographystyle{unsrt}
\bibliography{biblio2}
\end{document}